# A comparative study on magnetic order and field-induced magnetic transition in double perovskite iridates: $RE_2ZnIrO_6$ and $RE_2MgIrO_6$ (RE=Pr,Nd,Sm,Eu,Gd)

Yuxia Gao[1,2], Zhaoming Tian[1,2]* Longmeng Xu[1,2], Malik Ashtar [1,2], Zongtang Wan[1,2], Zhengcai Xia[1,2]*, Feng Yang[1,2], Songliu Yuan[1,2], Yuyan Han[3], Wei Tong[3]

[1] School of Physics, Huazhong University of Science and Technology, Wuhan 430074, P. R. China

[2] Wuhan National High Magnetic Field Center, Huazhong University of Science and Technology, Wuhan 430074, P. R. China

[3] Anhui Province Key Laboratory of Condensed Matter Physics at Extreme Conditions, High Magnetic Field Laboratory, Chinese Academy of Science, Hefei, 230031, PR China.

**Abstract:** We perform a comparative magnetic study on two series of rare-earth (RE) based double perovskite iridates $RE_2BIrO_6$ (RE=Pr,Nd,Sm-Gd;B=Zn,Mg), which show Mott insulating state with tunable charge energy gap from ~330 meV to ~560 meV by changing RE cations. For nonmagnetic RE=Eu cations, $Eu_2MgIrO_6$ shows antiferromagnetic (AFM) order and field-induced spin-flop transitions below Néel temperature ($T_N$) in comparison with the ferromagnetic (FM)-like behaviors of $Eu_2ZnIrO_6$ at low temperatures. For magnetic-moment-containing RE ions, $Gd_2BIrO_6$ show contrasting magnetic behaviors with FM-like transition (B=Zn) and AFM order (B=Mg), respectively. While, for RE=Pr, Nd and Sm ions, all members show AFM ground state and field-induced spin-flop transitions below $T_N$ irrespective of B=Zn or Mg cations. Moreover, two successive field-induced metamagnetic transitions are observed for $RE_2ZnIrO_6$ (RE=Pr,Nd) in high field up to 56 T, the resultant field temperature ($H$-$T$) phase diagrams are constructed. The diverse magnetic behaviors in $RE_2BIrO_6$ reveal that the $4f$-Ir exchange interactions between the RE and Ir sublattices can mediate their magnetism.

## 1. Introduction

The 5d iridates have attracted great attention as candidate materials for exploring exotic quantum phases of matter with unconventional electronic and magnetic ground states, such as novel Mott insulator [1], Kitaev spin liquid [2], Weyl semimetal [3,4] and even unconventional superconductors [5].The underlying mechanisms of these exotic phenomena are associated with strong spin-orbital coupling (SOC) and its interplay with large crystalline electric fields (CEF), electron correlation interactions (U) in comparable energy scales. As one prominent consequence of strong SOC in iridates, it leads to the formation of so-called novel "$j_{eff}$=1/2 Mott insulator" as initially discovered in $Sr_2IrO_4$ compound [6].Within this picture, the energy scale of competitive SOC, CEF and electron correlation interactions can be substantially tailored by local environment, symmetry or dimensionality [7,8], which can modify or even destroy the $j_{eff}$=1/2 state. In this respect, iridates with different lattice geometries or spin-exchange pathways provide template to explore diverse exotic quantum states. A variety of systems with different geometrical structure have been garnered considerable interest, such as the perovskite $Sr_2IrO_4$ and $SrIrO_3$ [6,9], honeycomb compounds $Na_2IrO_3$ and $Li_2IrO_3$ [10,11],hyperkagome $Na_4Ir_3O_8$ [12] and pyrochlore iridates $RE_2Ir_2O_7$(RE=rare earth) [13], etc.

* Author to whom correspondence should be addressed: tianzhaoming@hust.edu.cn
* xia9020@hust.edu.cn

Recently, the ordered double perovskite (DP) iridates with chemical formula $A_2BIrO_6$ (A is rare earth ions, B is 3d transition metal or alkaline earth metal) become intensively studied as a new system of " $j_{eff}$=1/2 Mott insulator" [14-16]. In this family, large space separated Ir ions displaying a "rock-salt" face-centered-cubic (FCC) symmetry, lets this system have new features beyond $ABO_3$ perovskites such as geometric spin frustration, weakened direct Heisenberg exchange, enhanced SOC effect or even emergence of Kitaev interactions[17,18]. In case of nonmagnetic A/B cations (A=La,Y; B=Zn,Mg), $A_2BIrO_6$ is a prototypical compound where magnetism is dictated by 5d electrons of Iridium. When A site is occupied by magnetic rare-earth (RE) cations (A=Pr,Nd,etc), the 4*f*-4*f* exchange interactions within RE ions at A sites exist, also the 4*f*-Ir exchange interactions between the A/B sublattices are possibly present. In this case, the 4*f* magnetic moments together with 4*f*-Ir interaction as additional term coupled to the $Ir^{4+}$ moments will enrich its magnetic and electronic behaviors, while the related issue is unclear.

Experimentally, the basic magnetic properties of $RE_2MgIrO_6$ (RE=La,Pr,Nd,Sm-Gd) compounds were previously characterized [19,20], most of them show antiferromagnetic (AFM) order. The recent studies reveal that the $RE_2BIrO_6$ (RE=La,Eu) with nonmagnetic A cations show disparate magnetic orders in respect to B=Mg or Zn [14,21,22], where $RE_2ZnIrO_6$ but not $RE_2MgIrO_6$ show ferromagnetic (FM)-like hysteresis below Curie temperature ($T_C$) [14,22]. While, as RE is extended to magnetic-moment-contained ions, their magnetic behaviors are still controversial, as example, the recent experimental studies on Sr-doped $Pr_2MgIrO_6$ show they are neither half-metallicity nor long-range ferrimagnetic ordering [23], sharply contrast to the theoretically predicted half-metallic antiferromagnetism/ferrimagnetism [24]. Moreover, very recent Raman scattering measurements reveal the existence of fractionalized excitations in $Sm_2ZnIrO_6$ [25], supporting its' possible spin liquid state. Thus, to clarify their controversial magnetic behaviors in $RE_2BIrO_6$, further studies are highly desirable.

Here, we report a comparative study on magnetic behaviors and field-induced magnetic transitions in two serial $RE_2BIrO_6$ (RE=Pr,Nd,Sm,Eu,Gd; B=Zn,Mg) double perovskite iridates, our electrical measurements show all members have Mott insulating states with tunable charge gap from ~330 meV to ~560 meV as changing RE from La to Gd. In case of RE=Eu,Gd, $RE_2MgIrO_6$ exhibit the AFM order and field induced spin flop transitions below Néel temperature ($T_N$) in comparison with the FM order with typical FM hysteresis of $RE_2ZnIrO_6$. As for RE=Pr, Nd and Sm, all compounds exhibit field induced spin-flop transitions below $T_N$. Moreover, two successive field induced metamagnetic transitions are observed for $RE_2ZnIrO_6$ (RE=Pr,Nd) under field up to 56 T, and the magnetic field temperature ($H$-$T$) phase diagrams are constructed.

## 2. Experiment

Two serial $RE_2BIrO_6$ (RE=La,Pr,Nd,Sm,Eu,Gd; B=Zn,Mg) polycrystals were synthesized by conventional solid state reaction method. The mixtures were prepared using rare-earth oxides (99.99% $RE_2O_3$ for RE =La,Nd,Sm,Eu and Gd; 99.9% $Pr_2O_3$) ZnO(99.9%), MgO (99.5%) and $IrO_2$(Tanaka Kikinzoku Kogyo) as starting materials. For the $RE_2BIrO_6$(RE=La, Pr,Nd,Sm) samples, stoichiometric mixtures were carefully ground and reacted at temperature of 1150°C in air for 4 days with intermediate grindings. For RE=Eu,Gd compounds, higher reaction temperature of 1200°C-1250°C is required to obtain pure phase sample. The attempts to synthesize the compounds with smaller size cations than RE=Tb by different heating procedures at ambient pressure were unsuccessful. Additionally, to analyze magnetic behaviors of $Eu_2BIrO_6$, isostructural $Eu_2BTiO_6$ compounds were also synthesized using similar heating sequences. The structure and phase purity were checked by the X-ray diffraction (XRD, Rigaku Smartlab) measurements with Cu K$\alpha$ radiation (λ=1.5418Å). To get the structural parameters, the XRD spectra were analyzed through Rietveld refinements using the material studio software.

Magnetic measurements were carried out by the superconducting quantum interference device magnetometer (SQUID, Quantum Design) and commercial Physical Property Measurement System (PPMS, Quantum Design) with a vibrating sample magnetometry (VSM) option in applied DC magnetic fields up to 14 T, including temperature dependent susceptibility from 2 K-300 K and isothermal field dependent magnetization $M$($H$) measurements. The electrical resistance was measured by a standard four-probe method. The pulsed field magnetizations up to 56 T were measured by the induction method at Wuhan National High Magnetic Field Centre (WHMFC) down to 2 K with a calibration by the DC magnetization data.

## 3. Results and Discussions

### 3.1 Structural analysis of $RE_2BIrO_6$

The double perovskite (DP) $RE_2BIrO_6$ crystallize into monoclinic structure with space group $P2_1/n$, as report in $RE_2MgIrO_6$ (RE=La-Nd,Sm-Gd) [19-20]. This space group allows for two crystallographically distinct octahedral sites in the double perovskite structure, the $BO_6$ and $IrO_6$ octahedras are ordered in a rock-salt manner and their tilting to accommodate the small size RE cations. As a representative, the crystal structure of $Pr_2ZnIrO_6$ is presented in figure 1(a), with Pr atoms at 4e(x y z) positions, Zn at 2c (0 0 0) positions, Ir at 2d (1/2 1/2 0) and three types of oxygen atoms at 4e(x y z) sites. Within the corner-shared connections of $ZnO_6$/$IrO_6$ octahedras, the $Pr^{3+}$ cations fit the distorted cubic network of these rock-salt lattices in the local coordination environment of $Ir^{4+}$/$Pr^{3+}$ atoms, as shown in figure 1(b,c). The $Pr^{3+}$ atom is coordinated by eight oxygen atoms, forming the irregular $Pr^{3+}$-O polyhedron with different $Pr^{3+}$-O bond lengths. Each $Ir^{4+}$ atom is surrounded by eight nearest-neighboring $Pr^{3+}$ atoms, the $Ir^{4+}$-O bond distance varies within an octahedron, where six Ir-O bond lengths are grouped into three values reflecting the octahedral distortions.

The powder X-ray diffraction (XRD) spectra are refined by Rietveld methods, which reveal all $RE_2BIrO_6$ (RE= La,Pr,Nd,Sm,Eu,Gd) compounds are single phase without detected phase impurities. The resultant structural parameters for $RE_2ZnIrO_6$ are shown in figure 2(a), where the lattice parameters *a* and *c* show regularly decrease with radius size of rare earth (RE) cations, in comparison with the slightly increase of *b* parameter. Similar trends for $RE_2MgIrO_6$ are also observed as report in Ref [19]. This behavior can be related to the tilting scheme of $ZnO_6$ and $IrO_6$ octahedra in the $P2_1/n$ space group, corresponding to the $a^-a^-b^+$ in the Glazer's notation [26]. In this tilting system, structural distortion driven by the reduction of $RE^{3+}$ will cause minimal modification of the *b* parameter. For the octahedral distortion in $RE_2BIrO_6$, it can be identified by the distorted B-O-Ir bond angle away from ideal 180°. The tilting of $(B,Ir)O_6$ octahedra is characterized by the average tilting angle defined as $\phi = (180 - \omega)/2$ ,where ω is the inter-octahedral B-O-Ir angle. In figure 2(b), the evolution of average tilting angle ($\phi$), the RE-O and Ir-O bond lengths are presented, it is noted that smaller RE-ion radius cause larger tilting angles from $La^{3+}$ to $Gd^{3+}$, namely larger monoclinic distortions. In terms of interatomic distances, the size of $RE^{3+}$ ions mainly affects the average <RE-O> bond lengths, which almost linearly correlate with the $RE^{3+}$ radius, but no systematic change of <Ir-O> average lengths are detected in these series.

### 3.2 Electrical transport of $RE_2BIrO_6$

Figures 3(a,b) show the temperature (*T*) dependence of electrical resistivity *ρ*(T) for $RE_2BIrO_6$ (B=Zn, Mg), respectively. For all compounds, a monotonic increase in $\rho$(T) as decreased $T$ reveals the insulating ground state. Among the families, the $\rho$(T) of compounds with smaller RE ionic radius show larger resistance over all temperatures. As shown in figures 3(c) and (d), the experimental data are fitted by the thermal activation model described by $\rho(T) = \rho_0 \exp(\Delta / 2k_B T)$ , where $\Delta$ is the activation energy and $k_B$ is the Bolzmann constant. At higher temperatures, the resistivity follows the thermally activation model, but it deviates as decreased temperatures. The estimated charge energy gaps (Δ) in temperature regions 270 K-350K are listed in Table 1, which varies from ~336 meV (RE=La) to ~517 meV (RE=Gd) for $RE_2ZnIrO_6$. The Mg-containing analogs display similar trend with relatively larger charge gaps. Additionally, the $\rho(T)$ data can be better described by a three dimensional (3D) Mott variable-range-hopping (VRH) model [27]: $\rho = \rho_0 \exp\left(\frac{T_0}{T}\right)^{1/4}$, where $\rho_0$ is the resistivity coefficient and $T_0$ is the localization temperature. This is indicated by the straight line fit of ln$\rho$ versus $T^{-1/4}$ in figures 3(e) and (f)]. The fitted parameters $T_0$ are given in Table 1. According to the VRH model, $T_0$ depends on the localization length $l$ and the density of states at the Fermi level $N$ ($E_F$) in a relation of $T_0 = 18 / k_B N\left(E_F\right) l^3$ [28]. Then, the electron hopping between the localized states can be responsible for the electrical conductivity, the increase of $T_0$ as changing RE from $La^{3+}$ to $Gd^{3+}$ can be related to the increased $l$. The 3D character in the electronic transport is reasonable since the building blocks are made of well-separated $IrO_6$ octahedrons with FCC-type arrangement of $Ir^{4+}$ lattice, the large space separated Ir ions and electron correlations may account for the charge localizations. Similar transport mechanisms were reported in other 5d DP oxides $Sr_2InReO_6$ and $Ba_2NaOsO_6$ [29,30]. In all, the $\rho$(T) curves reveal $RE_2BIrO_6$ have charge-gapped Mott insulating states, the gap size can be tuned in a range from ~330 meV to 560 meV by changing RE ions.

### 3.3 Magnetic behaviors of $RE_2BIrO_6$

The zero-field-cooled (ZFC) and field-cooled (FC) magnetic susceptibilities $\chi$(T) were measured under field $H$ = 0.1 T for $RE_2BIrO_6$, which are shown in figure 4(a)-(e), respectively. For RE=Pr,Nd and Sm, $\chi$(T) for both B=Zn and Mg compounds show antiferromagnetic (AFM) transition at Néel temperature $T_N$. Below $T_N$, no distinct divergences are observed between the ZFC and FC magnetizations. As shown in figure 4(f)-(h), the inverse susceptibility 1/$\chi$(T) at paramagnetic (PM) states are fitted by the Curie-Weiss (CW) Law: $\chi(T) = \frac{C}{T - \theta_{CW}}$ , where $C$ is the Curie constant and $\theta_{CW}$ is the Curie-Weiss temperature. For $Sm_2BIrO_6$, we fit 1/$\chi$(T) at low temperature region (30 K-70 K) since the $Sm^{3+}$ ions have high temperature Van-Vleck paramagnetic (PM) contributions [31], leading to its' deviation of linear dependence. The fitted $\theta_{CW}$ and effective magnetic moments ($\mu_{eff}$) are summarized in Table 2. Large negative $\theta_{CW}$ compared to $T_N$, as example, $\theta_{CW}$ = -53.3 K around 4 times large of $T_N$ for $Pr_2ZnIrO_6$, indicates the presence of magnetic frustration. The derived experimental moments are close to the theoretical value $\mu_{eff} = \sqrt{2\mu_{RE}^2 + \mu_{Ir}^2}$ with combinations of the moments of free $RE^{3+}$ ion $\mu_{RE} = g_J[J(J+1)]^{1/2}$ and $\mu_{Ir} = 2[1/2(1/2+1)]^{1/2}$ =1.73$\mu_B$/$Ir^{4+}$.

As shown in figure 4(d), $Eu_2ZnIrO_6$ shows ferromagnetic (FM)-like transition with $T_C$~11 K, below which the ZFC and FC magnetizations show large bifurcations. The FM-order state is further confirmed by the FM-like hysteresis below $T_C$ [see figure 5(a)]. In contrast, $Eu_2MgIrO_6$ shows AFM-type transition with $T_N$~10.4 K. Since the $Eu^{3+}$ ions ($^7F_0$,$4f^6$, S=3, $L$=3, $J$=0) possess no magnetic moment and the excited states $^7F_J$ ($J$=1,2,…) only give rise to weak temperature dependent Van-Vleck PM contribution [32], low temperature magnetic orders are dictated by the $Ir^{4+}$ moments in $Eu_2BIrO_6$. So, to quantatively evaluate their magnetic interactions, the Van-Vleck PM contribution ($\chi_{VV}$) from $Eu^{3+}$ ions should be subtracted [33]. Here, the susceptibility of isostructural $Eu_2BTiO_6$ (B=Zn,Mg) is used as $\chi_{VV}$ (see Appendix, figure 11), the yielded $1/\chi_{Ir} = 1/(\chi\text{-}\chi_{VV})$ is presented in figure 4(i). From the CW fitting, the obtained $\theta_{CW}$=9.4 K (Zn) and -15.3 K (Mg) reveal its dominant FM and AFM interactions, respectively.

To more clearly characterize the ferromagnetism in $Eu_2ZnIrO_6$, the Arrott analysis is used to determine the $T_C$ and ordered moment. As shown in figure 5(b), we show $(M)^{1/\beta}$ versus $(H/M)^{1/\gamma}$ plots. At 11 K, the isothermal magnetization has a linear dependence with $\beta$=0.65 and $\gamma$=1.21, which correct $T_C$=11 K. The saturated magnetization ($M_S$) at 2 K is determined to 0.37$\mu_B$/Ir from linear extrapolation of the straight line in modified Arrot plots. Under high field, its magnetization has linear field dependence without saturation up to 56 T (see figure 5(c)), this can be due to the dominant Van-Vleck PM contributions persistent up to high fields.

The isothermal magnetization $M(H)$ curves of $Eu_2MgIrO_6$ are shown in figure. 6(a). Below $T_N$, the $M(H)$ show pronounced steepening near critical field ($H_C$) featured a field-induced spin-flop transition, different from the FM hysteresis behaviors of $Eu_2ZnIrO_6$. Here, $H_C$ is defined by the peak position of derivative magnetization d$M$/d$H$, as shown in the figure 6(b). The $H_C$ decreases with increased temperatures towards zero at $T_N$. From the magnetic susceptibilities at different fields, the $\chi$(T) shows a peak at $T_N$ as $H$ < 3 T, further increased $H$, the peak becomes broadened and gradually smeared out. In this case, $T_N$ is defined by the peak of derivative d$M$/d$T$ as shown in the inset of figure 5(c). Based on above $M(H)$ and $\chi$(T) data, a schematic magnetic field/temperature ($H$-$T$) phase diagrams are constructed, shown in figure 6(d). The presence of spin flop reveals the evolution of magnetic structures from AFM to canted-antiferromagnetic ($C$-AFM) states.

The linear extrapolation of high field magnetization at 2 K yields net remnant magnetization ~0.27$\mu_B$/f.u., this indicates $Eu_2MgIrO_6$ have *C*-AFM phase with net FM moments above $H_C$.

As presented in figure 4(e), $Gd_2BIrO_6$ (B=Zn,Mg) show distinct magnetic behaviors, $Gd_2ZnIrO_6$ shows FM-like behavior supported by the steep increase of FC magnetization near $T_C$~22 K and well-shaped FM hysteresis below $T_C$ (See figure 7(a)). While, $Gd_2MgIrO_6$ exhibits the AFM-like transition with $T_N$~20.8 K, where $T_N$ is defined by the peak of derivative d*M*/d*T* [see the inset of figure 7(c)]. Moreover, different sign of $\theta_{CW}$ with $\theta_{CW}$=1.5 K (B=Zn) and $\theta_{CW}$=-5.3 K (B=Mg) also support the different dominant FM or AFM interactions, respectively. For $Gd_2ZnIrO_6$, the remnant magnetization $M_r$=2.52$\mu_B$/f.u. at 2 K is much larger than the maximum value $M_S$=1.0$\mu_B$/$Ir^{4+}$, this imply that the $Gd^{3+}$ moments have FM contributions. As a comparison, this typical FM behaviors are contrast to magnetic behaviors reported quite recently in $Gd_2ZnIrO_6$, where it exhibits wasp-waist-like hysteresis and small remnant magnetization ($M_r$~1.5 $\mu_B$/f.u.) [22]. As shown in the inset of figure 7(a), its magnetization saturates at ~ 20 T with $M_S$=14.8$\mu_B$/f.u, this value is consistent with the theoretical saturated moments $=(2g_J J_{Gd}+g_J J_{Ir})\mu_B/\text{f.u.}=(2\times2\times\frac{7}{2}+2\times\frac{1}{2})\mu_B/\text{f.u.}=15\mu_B/\text{f.u.}$ for combinations of $2Gd^{3+}$ and $Ir^{4+}$ moments. For $Gd_2MgIrO_6$, *M*(*H*) curves show waist-restricted shape hysteresis with nearly zero remnant magnetization. This unusual hysteresis reveal the field induced spin-flop transition from the magnetic reorientation of $Gd^{3+}$ and $Ir^{4+}$ sublattices, the resultant *H*-*T* phase diagrams are shown in figure 6(d). Compared to $Eu_2MgIrO_6$, the smaller $H_C$~0.8 T at 2 K can be related to the large gained Zeeman energy of $Gd^{3+}$ spins coupled to external fields. Moreover, since the $Gd^{3+}$ moments ( $7.94\mu_B/\text{Gd}^{3+}$ ) is much larger than $Ir^{4+}$ moments ( $1.73\mu_B/\text{Ir}^{4+}$ ), the monotonically increased magnetization at high fields in $Gd_2BIrO_6$ can be attributed to the alignment of $Gd^{3+}$ moments.

For $RE_2BIrO_6$ (RE=Pr,Nd, Sm; B=Zn,Mg), the isothermal DC magnetization curves at 2 K are shown in figure 8(a),(c) and (e), respectively. At low fields, linear dependent magnetization behaviors are in consistent with their AFM ground states. Further increased field, a field induced spin-flop transitions appear, the small hysteresis indicates the characteristic of 1$^{st}$ order transition, as also seen from the d*M*/d*H* curves in the inset of figure 8(a). To further identify the field-induced metamagnetic transitions, pulsed field magnetization measurements up to 56 T are shown in figure 8(b),(d) and (f). As seen, the 2$^{nd}$ metamagnetic transition occurs for $RE_2ZnIrO_6$ (RE=Pr,Nd), where another anomaly of d*M*/d*H* appears near $H_{c2}$ [see the inset of figure 8(b) and (d)]. While for Mg-analogues, no 2$^{nd}$ field-induced transitions are detected. Here, the observed metamagnetic transitions reveal the existence of spin-reorientation of the $Pr^{3+}$/$Nd^{3+}$ and $Ir^{4+}$ moments. Additionally, the saturated magnetization moment ($M_S$) at 56 T is far below the theoretical saturation moment value $M_S=2M_{RE}+M_{Ir}$, as example, the experimental $M_S$~3.49$\mu_B$/f.u for $Pr_2ZnIrO_6$ is quite smaller than expected $M_S=(2g_J J+1)\mu_B$/f.u.=$(2\times0.8\times4+1)\mu_B$/f.u.= 7.4$\mu_B$/f.u. This can be related to the reduced moments of $Pr^{3+}$/$Nd^{3+}$ at low temperatures due to the low-lying CEF effects [34,35]. Further taking into account its Ising-like magnetic anisotropy, the saturated magnetization per $Pr^{3+}$ ions should be $g_J J/2$ ~1.15-1.3$\mu_B$, then the yielded $M_S=(2\times\frac{1}{2}g_J J+1)\mu_B\sim3.3\text{-}3.6\mu_B$ agrees well with our experimental value. Here, the used moment of $Pr^{3+}$ is consistent with the detected moment value (2.3$\mu_B$) by neutron powder diffraction in the isostructural $Pr_2LiRuO_6$ system [36]. For $Sm_2ZnIrO_6$, high-field magnetizations show no sign of saturation up to 56 T, its maxima moment ~1.2$\mu_B$/*f.u.* is far from saturated value in respect to effective moments ~0.85$\mu_B$/$Sm^{3+}$ and 1.73$\mu_B$/$Ir^{4+}$. Thus, no 2nd metamagnetic transition is detected possibly because the $H_{c2}$ is beyond our measured field regions.

For $Pr_2ZnIrO_6$, the isothermal magnetization curves for field sweep-up branches under DC and pulsed magnetic fields are shown in figure 9(a) and (c), field-induced two successive metamagnetic transitions are detected below $T_N$. The evaluated $H_{C1}$ from the peak position of d*M*/d*H* curves decreases as increased temperatures (see figure 9(b)). In figure 9(d), we present the magnetic susceptibilities measured at different fields. $T_N$ is defined by the peak of derivative d*M*/d*T*, as shown in the inset of figure 9(d). Using the $H_C$ and $T_N$ obtained from above *M*(*H*) and *M*(T) data, the *H*-T phase diagram of $Pr_2ZnIrO_6$ is constructed, as shown in figure 10(a). Similar *H*-T phase diagram of $Nd_2ZnIrO_6$ is also mapped based on high field magnetization curves (see Appendix, figure 12), as shown in figure 10(b). At low temperatures, The $H_{C1}$ for field sweep-up and field sweep-down are different as shown in figure 8, as characteristics of 1st order transition. Above 7 K, the magnetization for field sweep-up and sweep-down completely overlaps. To better understand this transition, we do a linear magnetization extrapolation as show in in figure 8(a),(c) and (f). Considering that the saturated moment ~0.27-0.37 $\mu_B$/$Ir^{4+}$ in $A_2ZnIrO_6$ (A=La,Eu) [14,37], different yielded magnetization values of 0.92 $\mu_B$/f.u.(A=Pr),1.12$\mu_B$/f.u. (Nd) and 0.17$\mu_B$/f.u (A=Sm) reveal its origin from the spin-reorientation of $RE^{3+}$ moments. Above $H_{C1}$, the $RE^{3+}$ sublattices form a *C*-AFM ordered state with net FM components. Additionally, the 2nd metamagnetic transition observed at high field for B=Zn instead of Mg-analogues reveal the existence of different magnetic configurations for Zn/Mg series at high fields.

## 3.4 Discussions

Among the DP iridates, magnetic behaviors of $A_2BIrO_6$ (A=La,Eu) are dictated by the $Ir^{4+}$ ( $j_{eff}$=1/2) moment on the face-centered-cubic (FCC) lattices. From the structural analysis, the $IrO_6$ octahedron in $A_2BIrO_6$ series has a rotation $\phi$ relative to *c* axis. In strong spin-orbital coupled (SOC) limit, Ir moments can be rigidly locked to the octahedral rotation. In this case, this octahedral rotations can induce a collective deviation of $j_{eff}$=1/2 moment canted away from [110] axis in the *ab* plane, as revealed by recent neutron powder diffractions and theoretical analysis in $La_2BIrO_6$ [17,18], where the $Ir^{4+}$ moments form A-type AFM ground state, but have different magnetic configurations in respect to B=$Zn^{2+}$ or $Mg^{2+}$ ions. In $La_2ZnIrO_6$, the $Ir^{4+}$ moments prefer FM coupling within the *ab* plane and staggered AFM between the adjacent *ab* layers [17]. In this spin arrangement, net FM moments can arise from the isospin canting effect determined by the staggered rotation angle ($\phi$), similar spin configurations should also be formed in $Eu_2ZnIrO_6$ responsible for appearance of net FM moments. While, in $A_2MgIrO_6$ (A=La,Eu), the FM planes are in the xz planes and stack AFM along y axis, so no net FM moment is obtained. The observed spin flops in $Eu_2MgIrO_6$ reflect the evolution of magnetic structures from AFM to *C*-AFM state, above $H_C$, the *C*-AFM phase has net FM moments.

For compounds containing magnetic 4*f* RE ions, the introduced RE ions can dramatically alter their magnetic behaviors. For larger size RE cations (RE=Pr, Nd and Sm), all members of two branches show AFM ground state and field induced metamagnetism, but $Gd_2BIrO_6$ (B=Mg,Zn) resemble magnetic features as observed in $A_2BIrO_6$ (A=La,Eu) analogs with FM (B=Zn) and AFM (B=Mg) ordered states, respectively. The above diverse magnetic behaviors cannot solely be attributed to the exchange coupling of $Ir^{4+}$ sublattices, the intra-sublattice RE-RE exchange interaction and inter-sublattice 4f-Ir exchange couplings should be included. Among $RE^{3+}$ ions, $Gd^{3+}$ is special and has half-filled 4f shell ($4f^7$, S=7/2, *L*=0), which usually show quasi-isotropic magnetic interactions due to the absence of CEF effect. This is relatively simple compared to other $RE^{3+}$-based systems, where magnetic anisotropy is also affected by the CEF splitting. Given that the absence of Gd-Ir magnetic couplings, the Gd-Gd magnetic correlations should occur at much lower temperatures ~ 1 K [38]. Here, $Gd_2BIrO_6$ show highest transition temperatures (~20 K) accompanied by enhanced magnetizations, indicative of the ordering of $Gd^{3+}$ magnetic sublattices below $T_N$ or $T_C$ via the Gd-Ir coupling. The Gd-Ir exchange interactions can be produced by the four first-neighbor $Ir^{4+}$ of each $Gd^{3+}$ site, and the $Gd^{3+}$ spins show some similar features of $Ir^{4+}$ sublattices with compatibility. This can explain the resemblance of magnetic behaviors in $Gd_2BIrO_6$ compared to its $Eu_2BIrO_6$ counterparts. $Gd_2ZnIrO_6$ has FM hysteresis with larger FM component $M_S$=4.2(2)$\mu_B$/f.u. at 2 K, ~28% of its saturated moments ($M_S$=15$\mu_B$/f.u.), signify the ferromagnetically coupled $Gd^{3+}$ and $Ir^{4+}$ sublattices.

As for RE=Pr, Nd and Sm, all members of $RE_2BIrO_6$ share AFM ground state without FM components irrespective of B=Zn or Mg, this is sharply contrast to the Gd counterpart. Moreover, the large negative $\theta_{CW}$ with a moderate frustrated factor, signify the presence of short range AFM correlation above $T_N$. Here, it should be noted that this AFM order can't be ascribed to either the RE-RE coupling or Ir-Ir interactions along, since the RE-RE interactions alone are weak, as example, the isostructural $RE_2LiIrO_6$ show no magnetic order down to 2 K [39]. Also, the dominant Ir-Ir interactions alone will lead to distinct FM (B=Zn) and AFM (B=Mg)-like behaviors as observed in the RE=La and Eu systems. In this respect, the incorporated RE sublattices seem to destroy the delicate balance of different Ir-Ir exchange interactions for B=Mg or Zn, respectively. The observed metamagnetic transitions in $RE_2BTiO_6$ (RE=Pr,Nd, Sm) correspond to the spin-flop transition from AFM to *C*-AFM state. For $Nd_2ZnIrO_6$, quite recent neutron diffraction reveals $Nd^{3+}$ moments have AFM state, where magnetic propagation vector along (1/2 1/2 0) at zero field [22]. As field above $H_{C1}$, further increased magnetizations favor the *C*-AFM state. This is consistent with canted antiferromagnetism, where the increased spin-tilting along field direction happens with increased fields. So, the 1$^{st}$ metamagnetic transition should originate from the spin-flip of $RE^{3+}$ sublattices. For $RE_2ZnTiO_6$ (RE=Pr,Nd), the 2$^{nd}$ metamagnetic transitions reveals another spin-reorientation of magnetic sublattices, then become the full polarized FM state (PS) at high fields ($H$> 40 T), as illustrated in the $H$-$T$ phase diagrams.

Compared with spin-only $Gd^{3+}$ ions, the main difference for RE=Pr,Nd and Sm is the presence of orbital moments and substantial CEF, which can tune the exchange interactions between $Ir^{4+}$ and $RE^{3+}$ moments in the exchange path by orbital hybridization. Additionally, three interactions including intra-sublattice Ir-Ir, RE-RE interaction and inter-sublattice RE-Ir exchange coupling are coexistent, all of them can play a role but the relative strength and sign of above interactions can strongly affect its' magnetic behaviors. Similar phenomena have been well reported in RE-based pyrochlore iridates, where the 4f-Ir coupling can dramatically tune their magnetic and electronical properties [40-42]. So, the RE-Ir exchange interactions in RE-based DP iridates provide alternative route to produce new emergent magnetic phases.

# 4. Conclusions

We performed the comparative study of magnetic order and field-induced metamagnetic transitions on two serial RE-based DP iridates $RE_2BIrO_6$ (A=Pr,Nd,Sm-Gd;B=Zn,Mg), which show Mott insulating state with charge energy gap from 330 meV to 560 meV with the variation of RE ions. In case of nonmagnetic RE=Eu cations, $Eu_2MgIrO_6$ exhibits AFM ground state and a field-induced spin flop transition from AFM to *C*-AFM states below $T_{N,}$ in contrast to the FM behaviors of $Eu_2ZnIrO_6$. For magnetic-moment-containing RE cations, the AFM-ordered $Gd_2MgIrO_6$ show field-induced spin-flop transition compared to the well-shaped FM hysteresis of $Gd_2ZnIrO_6$ below $T_C$. While, for RE=Pr, Nd and Sm, both Zn- and Mg-branches share similar AFM magnetic ground state and field induced metamagnetic transitions. Finally, two successive field-induced metamagnetic transitions are detected in high magnetic field up to 56 T and resultant magnetic phase diagram are constructed for both $Pr_2ZnIrO_6$ and $Nd_2ZnIrO_6$ compounds.

During the writing of the present manuscript, we became aware of another study investigating the magnetic properties of serial $RE_2ZnIrO_6$(RE=Nd,Sm,Eu,Gd) compounds conducted by Vogl M *et al*. [22] Based on high field magnetization measurements with field to ~56 T, our work showed the distinct magnetic behaviors and field-induced magnetic transitions in two serial $RE_2BIrO_6$ compounds. This also allowed us to construct the *H*-T phase diagrams for the metamagnetic transitions in $RE_2ZnIrO_6$ (RE=Pr,Nd). We have compared the present magnetic behaviors with their results in the main context.

## Conflicts of interest

There are no conflicts to declare.

### Acknowledgements

We acknowledge financial support by the National Natural Science Foundation of China (Grant Nos. 11874158 and 11674115), and by fundamental research funds for the central universities (Grant No. 2018KFYYXJJ038 and 2019KFYXKJC008). We would like to thank the staff of the analysis center of Huazhong University of Science and Technology for their assistance in various measurements.

## Appendix

### 1. The analysis of magnetic susceptibilities of $Eu_2BIrO_6$ (B=Zn,Mg)

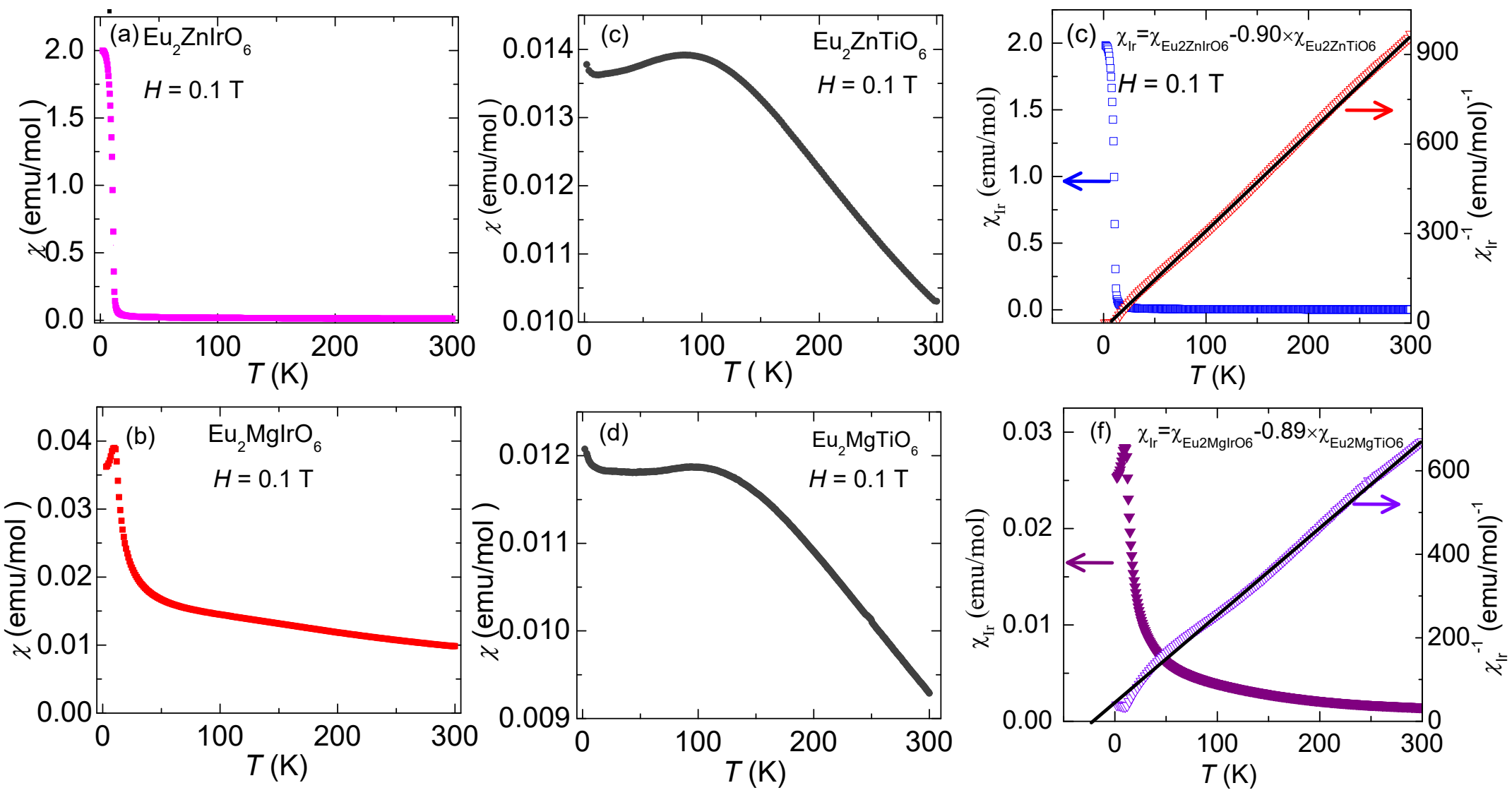

Figure 11. Temperature dependence of magnetic susceptibilities under *H*=0.1 T for (a) $Eu_2ZnIrO_6$,(b) $Eu_2MgIrO_6$,(c) $Eu_2ZnTiO_6$ and (d) $Eu_2MgTiO_6$, respectively. The resultant magnetic susceptibilities ($\chi_{Ir}$) and inverse magnetic susceptibilities ($1/\chi_{Ir}$) for (e) $Eu_2ZnIrO_6$ (f) $Eu_2MgIrO_6$, the black line shows the linear Curie-Weiss fitting.

Temperature dependence of magnetic susceptibility of $Eu_2BIrO_6$(B=Zn,Mg) under $H$=0.1 T are shown in figure 11(a,b). At high temperatures, the inverse susceptibility violates the Curie-Weiss law due to the presence of Van-Vleck paramagnetic (PM) contributions from $Eu^{3+}$($L$=3,$S$=3,$J$=0) ions [32,33], where electron occupations at excited states can give temperature-dependent contribution to susceptibility at high-temperatures. In this case, high temperature $\chi$(T) for $Eu_2BIrO_6$ should include two PM components from both $Ir^{4+}$ and $Eu^{3+}$ sublattices, as shown in figure 11(a) and (b). In this case, we analyze it by a modified Curie-Weiss Law as below: $1/\chi = 1/(\chi_{\mathrm{Ir}} + \chi_{VV})$ , where $\chi_{\mathrm{Ir}}$ and $\chi_{\mathrm{VV}}$ are magnetic components from $Ir^{4+}$ and $Eu^{3+}$ sublattices, respectively. To evaluate the magnetic interactions between $Ir^{4+}$ moments, the Van-Vleck PM part ($\chi_{\mathrm{VV}}$) is subtracted from the $\chi$(T) by using the susceptibility of $Eu_2BTiO_6$ (B=Zn,Mg) [ see Figure 11(c) and (d)]. Here,$Eu_2BTiO_6$ have isostructural to the $Eu_2BIrO_6$. The resultant $\chi_{\mathrm{Ir}}$ and $1/\chi_{\mathrm{Ir}} = 1/(\chi\text{-}\chi_{VV})$ are shown in figure 11(e,f), respectively. From high temperature (180K–300 K) Curie-Weiss fitting, the obtained $\theta_{\mathrm{CW}}$=9.4 K (Zn) and $\theta_{\mathrm{CW}}$=-15.3 K (Mg) reveal its dominant FM and AFM interactions, respectively

## 2. High field magnetizations of $Nd_2ZnIrO_6$

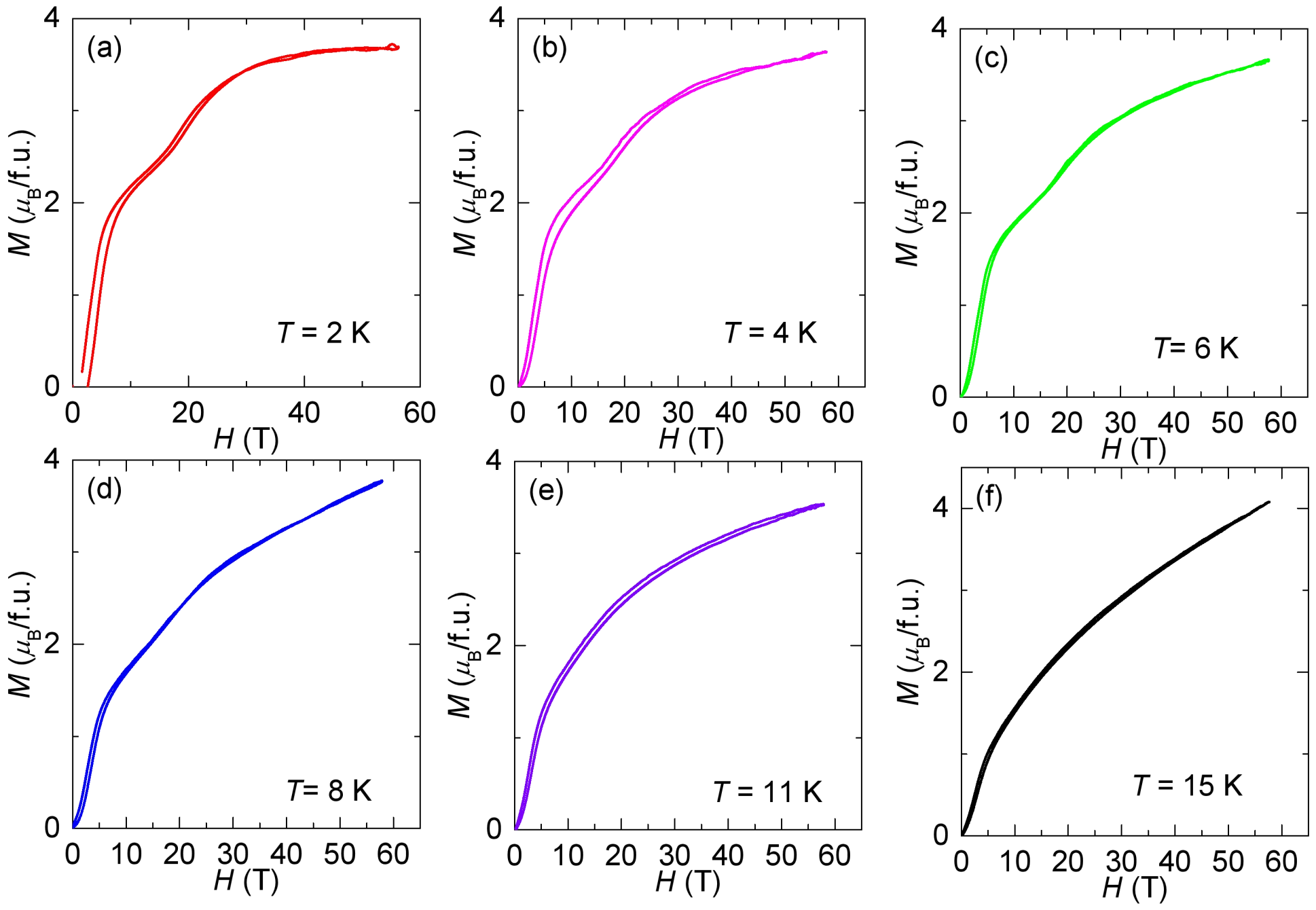

Figure 12. The field dependence of magnetizations with field up to 56 T at different temperatures for $Nd_2ZnIrO_6$.

The field dependent magnetization curves of $Nd_2ZnIrO_6$ are measured at different temperatures from 2 K to 15 K, the selected $M$($H$) curves are shown in figure 12. Below $T_N$, *the* field dependent magnetizations show two successive metamagnetic transitions, where the critical field ($H_{C2}$) is determined from the peak position of derivative magnetization

d*M*/d*H*. Combined the *M*(*H*) and *M*(*T*) curves, the $H_{C2}$ is used to construct the field-temperature *H*-T phase diagram of $Nd_2ZnIrO_6$.

## List of Tables

**Table 1.** The activation energy (△) fitted by the thermal activation model, and character temperature ($T_0$) by the VRH model for $RE_2BIrO_6$ (RE=La,Pr,Nd,Sm-Gd)

| | La | Pr | Nd | Sm | Eu | Gd |
|---|---|---|---|---|---|---|
| **△(meV)** | | | | | | |
| $RE_2ZnIrO_6$ | 336 | 386 | 417 | 425 | 460 | 517 |
| $RE_2MgIrO_6$ | 382 | 388 | 415 | 462 | 518 | 562 |
| **$T_0^{1/4}$(K$^{1/4}$)** | | | | | | |
| $RE_2ZnIrO_6$ | 126 | 131 | 137 | 139 | 153 | 166 |
| $RE_2MgIrO_6$ | 127 | 124 | 137 | 146 | 161 | 173 |

**Table 2.** The magnetic parameters of $RE_2BIrO_6$ including magnetic ordering temperatures ($T_C$,$T_N$), Curie-Weiss temperature($\theta_{CW}$) and effective magnetic moment ($\mu_{eff}$). The Curie-Weiss fitting is taken in temperature range (30 K-70 K) for $Sm_2BIrO_6$ and in temperature range (180 K -300 K) for $RE_2BIrO_6$ (RE=La,Pr,Nd, Eu, Gd; B=Zn,Mg).

| $RE_2BIrO_6$ | $T_N$,$T_C$(K) | $\theta_{cw}$(K) | $\mu_{eff}$($\mu_B$/f.u.) | $\mu_{fi}=\sqrt{2\mu_{RE}^2+\mu_{Ir}^2}$ ($\mu_B$/f.u.) |
|---|---|---|---|---|
| **B=Zn** | | | | |
| RE=La | 8.4 | -5.6 | 1.45 | 1.73 |
| RE=Pr | 12.0 | -53.3 | 5.52 | 5.36 [a] |
| RE=Nd | 15.4 | -53.8 | 5.54 | 5.41[a] |
| RE=Sm | 13.2 | -29.0 | 2.32 | 2.12[b] |
| RE=Eu | 11.0 | 9.4 | 1.51 | 1.73 |
| RE=Gd | 22.2 | 1.05 | 11.35 | 11.35[a] |
| **B=Mg** | | | | |
| RE=La | 11.4 | -19.0 | 1.52 | 1.73 |
| RE=Pr | 14.0 | -49.0 | 5.42 | 5.36[a] |
| RE=Nd | 11.8 | -51.3 | 5.32 | 5.41[a] |
| RE=Sm | 14.6 | -48 | 2.25 | 2.12[b] |
| RE=Eu | 10.8 | -15.3 | 1.53 | 1.73 |
| RE=Gd | 20.8 | -5.4 | 11.35 | 11.35[a] |

[a] $\mu_{RE}$ ($\mu_B$) and $\mu_{Ir}$ ($\mu_B$) are magnetic moments calculated by $g[J(J+1)]^{1/2}$ for free $RE^{3+}$ ions and ideal $j_{eff}$=1/2 $Ir^{4+}$ ions, [b] $\mu_{RE}$ ($\mu_B$) is the Van Vleck PM moment (0.85$\mu_B$) per $Sm^{3+}$ ion and $\mu_{Ir}$ ($\mu_B$) is moment per $Ir^{4+}$ ion.

## Figure captions

**Figure 1**. (a) Crystal structure of double perovskite $Pr_2ZnIrO_6$, in which $IrO_6$ and $ZnO_6$ are drawn as octahedra. Dark cyan, orange and gray spheres correspond to Pr, Ir and Zn sites. (b,c) The local coordinated environments of magnetic $Pr^{3+}$ and $Ir^{4+}$ ions in the connection of distorted $IrO_6$ octahedra and $PrO_8$ polyhedron.

**Figure 2**. (a) The variation of Lattice parameters for $RE_2ZnIrO_6$ as a function of $RE^{3+}$ ionic radius, (b) variation of RE-O/Ir-O bond lengths and tilting angle ($\phi$) for $RE_2ZnIrO_6$ with ionic radius of $RE^{3+}$.

**Figure 3**. Temperature dependence of resistivity and fitting data for $RE_2BIrO_6$ series. (a,b) $\rho$ versus $T$, (c,d) ln$\rho$ versus $T$, the solid lines are fitted by thermal activation model. (e,f) ln$\rho$ versus $T^{-1/4}$, the solid lines represent the fitting by VRH model.

**Figure** 4. (a-e) The ZFC and FC magnetic susceptibilities under $H$=0.1 T for $RE_2BIrO_6$(RE=Pr,Nd,Sm,Eu,Gd), (f-j) the inverse susceptibilities of FC magnetizations for $RE_2BIrO_6$, the black line shows the Curie-Weiss fitting. The inverse susceptibilities for $Eu_2BIrO_6$ are subtracted the Van Vleck PM parts.

**Figure 5**. (a) The isothermal $M(H)$ curves and (b) modified Arrot plot $(M)^{1/\beta}$ versus $(H/M)^{1/\gamma}$ for initial magnetizations of $Eu_2ZnIrO_6$, (c) Field dependent magnetization curves under pulsed magnetic fields at $T$= 2 K for $Eu_2ZnIrO_6$, the magnetization value is calibrated by the DC magnetization data measured by SQUID.

**Figure 6**. (a) The isothermal $M(H)$ curves of $Eu_2MgIrO_6$, the dashed red line represents the linear extrapolation of high-field magnetizations (b) the derivative magnetization d$M$/d$H$ curves for $Eu_2MgIrO_6$,(c) $M$(T) curves at different fields for $Eu_2MgIrO_6$, the inset shows the d$M$/d$T$ curves with $H$= 5 T. (d) the constructed $H$-$T$ phase diagram based on the $M(H)$ and $M(T)$ data.

**Figure 7.** The isothermal $M(H)$ curves at different temperatures for (a) $Gd_2ZnIrO_6$ and (b) $Gd_2MgTiO_6$, respectively. The inset in (a) shows high field magnetizations at $T$= 2 K for $Gd_2ZnIrO_6$, inset in (b) shows the d$M$/d$H$ curves for $Gd_2MgTiO_6$, (c) $M$(T) curves at different fields for $Gd_2MgIrO_6$, the inset shows the d$M$/d$T$ curves at 1 T, (d) the $H$-$T$ phase diagram based on the $M(H)$ and $M(T)$ data.

**Figure 8**. The DC magnetization and pulsed field magnetization curves at 2 K for (a,b) $Pr_2BIrO_6$, (c,d) $Nd_2BIrO_6$ and (e,f) $Sm_2BIrO_6$, respectively. The dashed black lines in (a,c,f) show the linear magnetization extrapolations, the inset in (a,b,d) show the derivative magnetization d$M$/d$H$ curves.

**Figure 9**. (a) The DC magnetization and (b) derivative magnetization d$M$/d$H$ curves for $Pr_2ZnIrO_6$. (c) The pulsed field magnetization curves of $Pr_2ZnIrO_6$, inset shows the d$M$/d$H$ curves (d) $M$(T) curves at different fields for $Pr_2ZnIrO_6$, the inset shows the d$M$/d$T$ curves

under *H*= 7 T.

**Figure 10.** The constructed *H-T* phase diagrams based on the *M*(*H*) and *M*(*T*) data for (a) $Pr_2ZnIrO_6$ and (b) $Nd_2ZnIrO_6$. The diagram is composed of antiferromagnetic (AFM), canted-antiferromagnetic (*C*-AFM), polarized ferromagnetic state (PS) and paramagnetic (PM) regions.

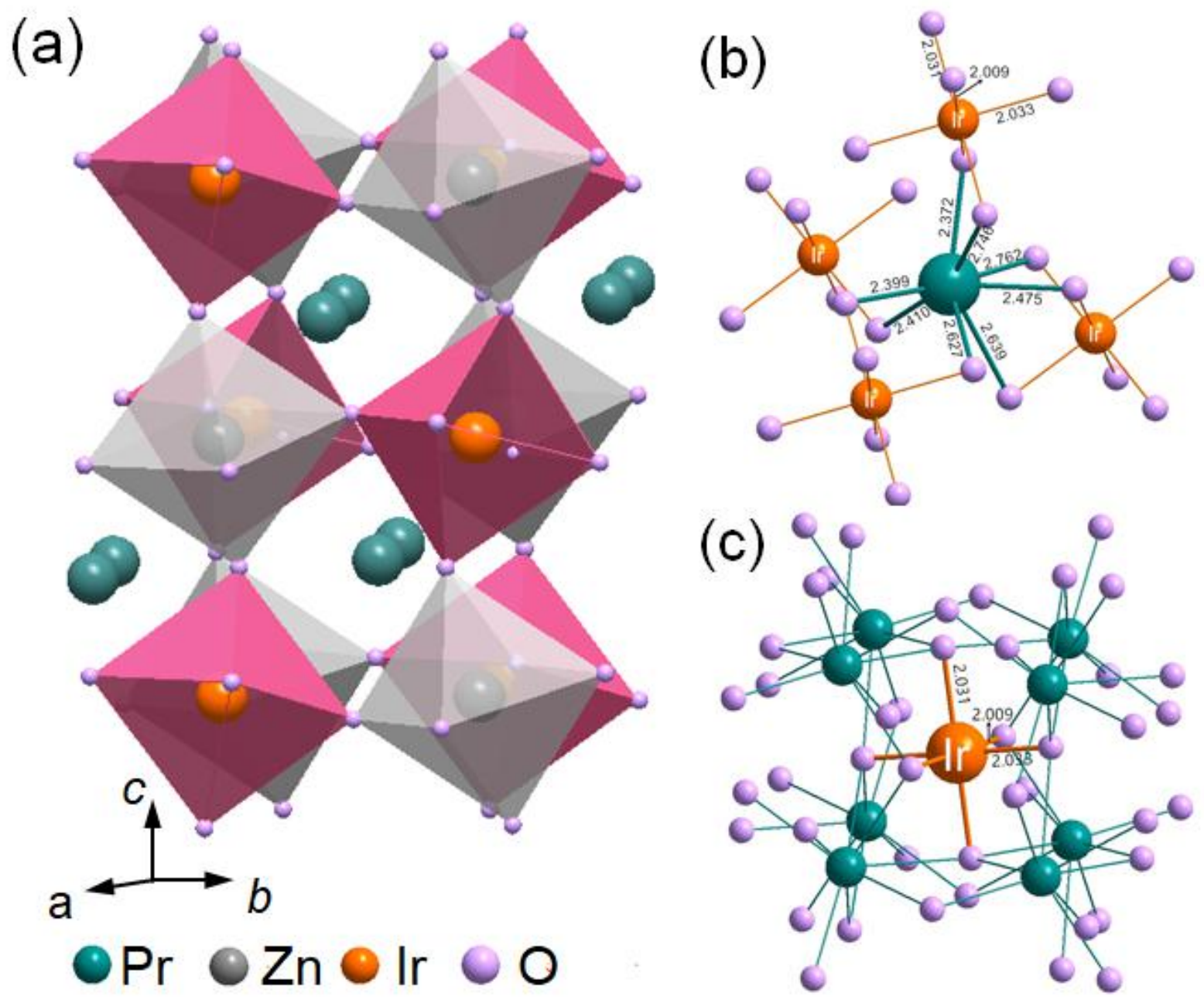

Figure 1

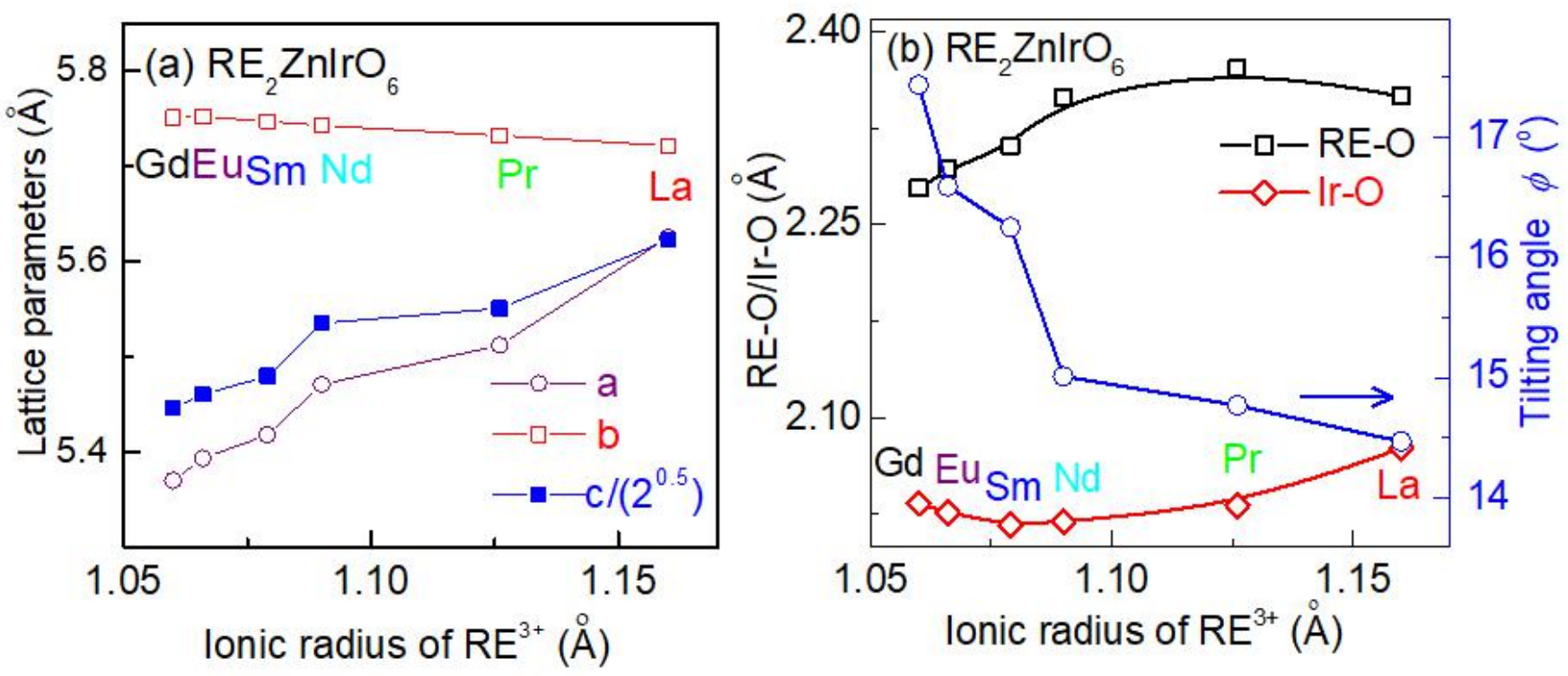

Figure 2

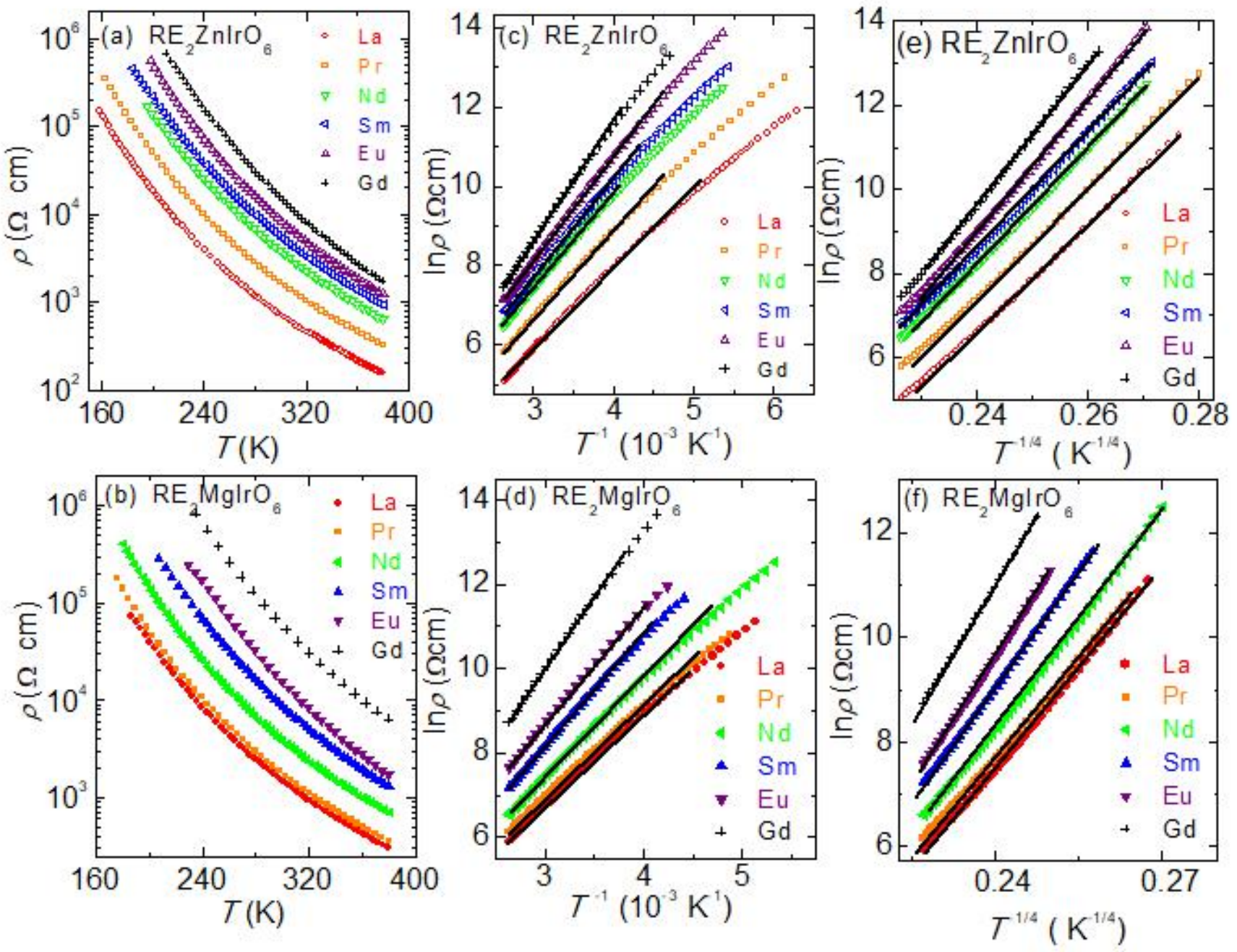

Figure 3

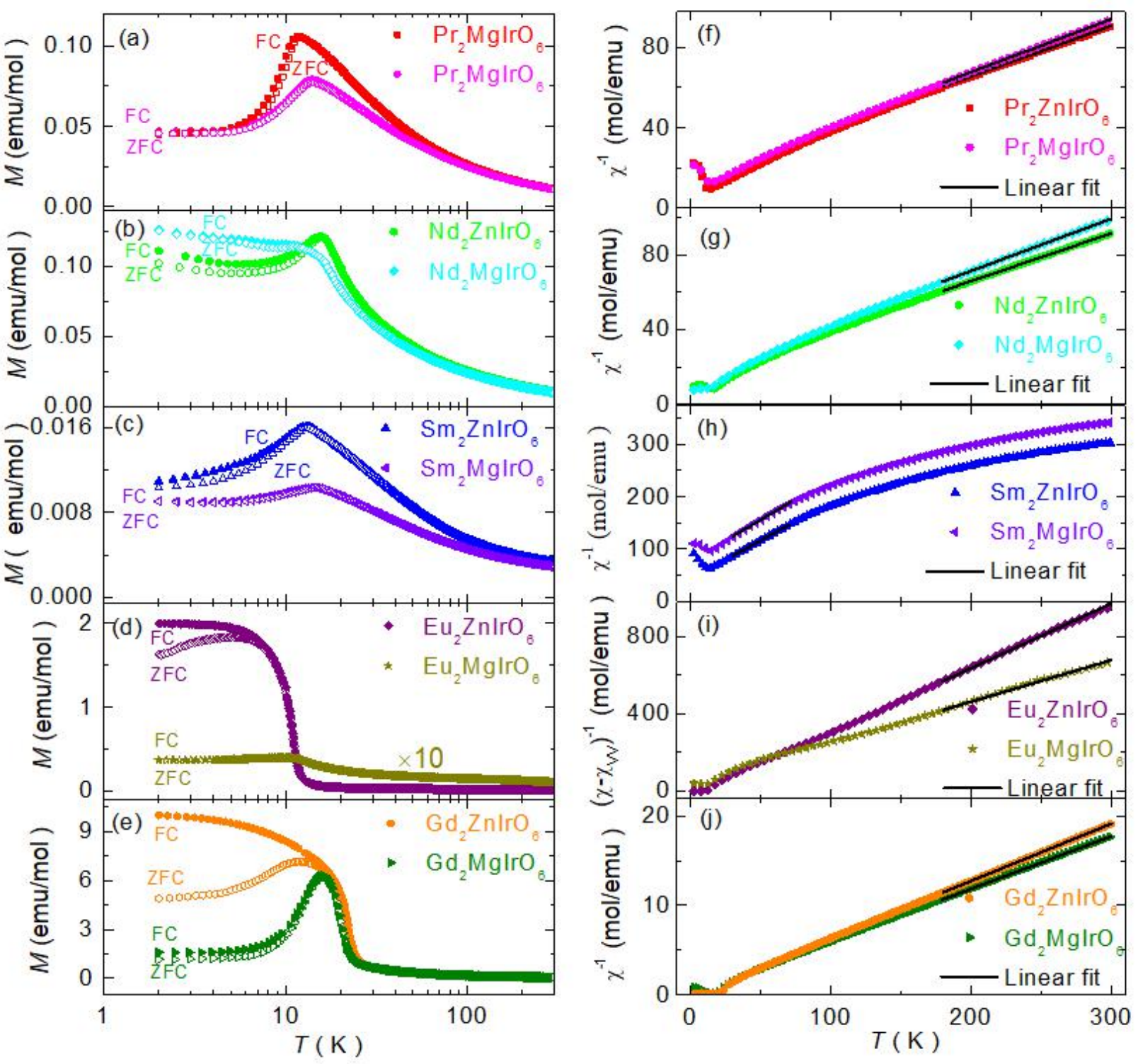

Figure 4

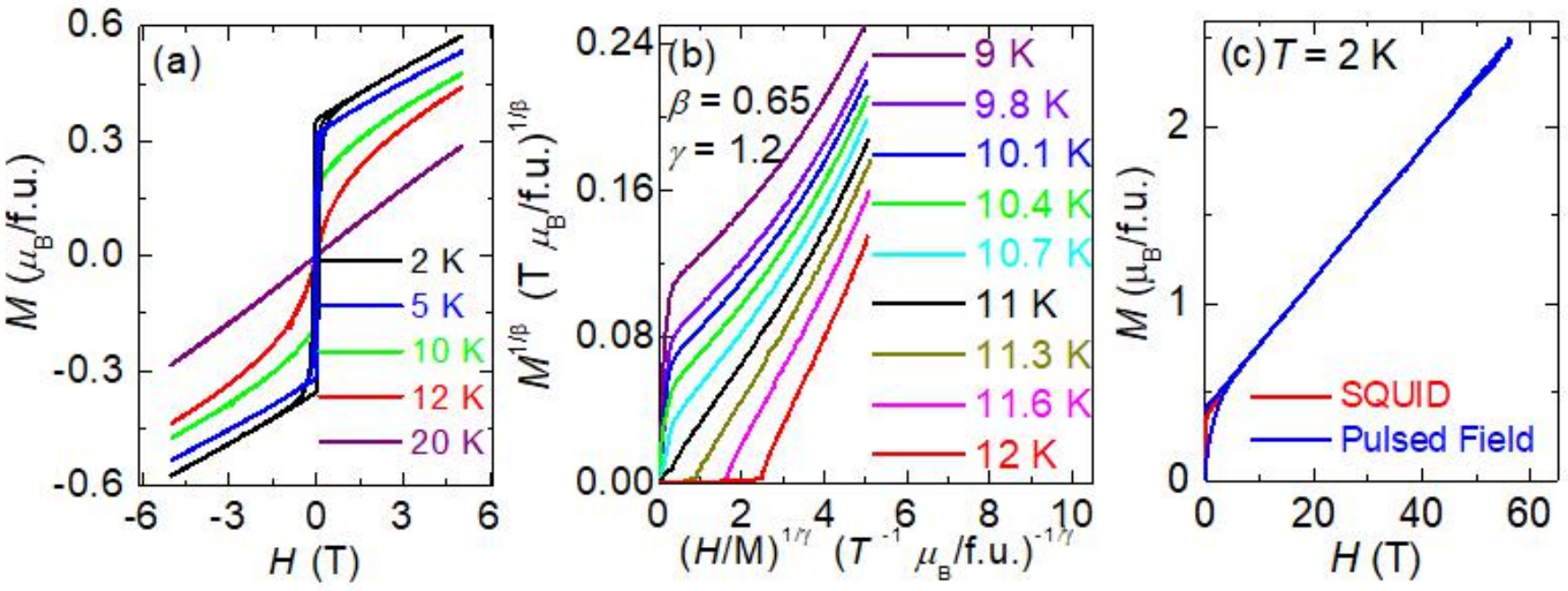

Figure 5

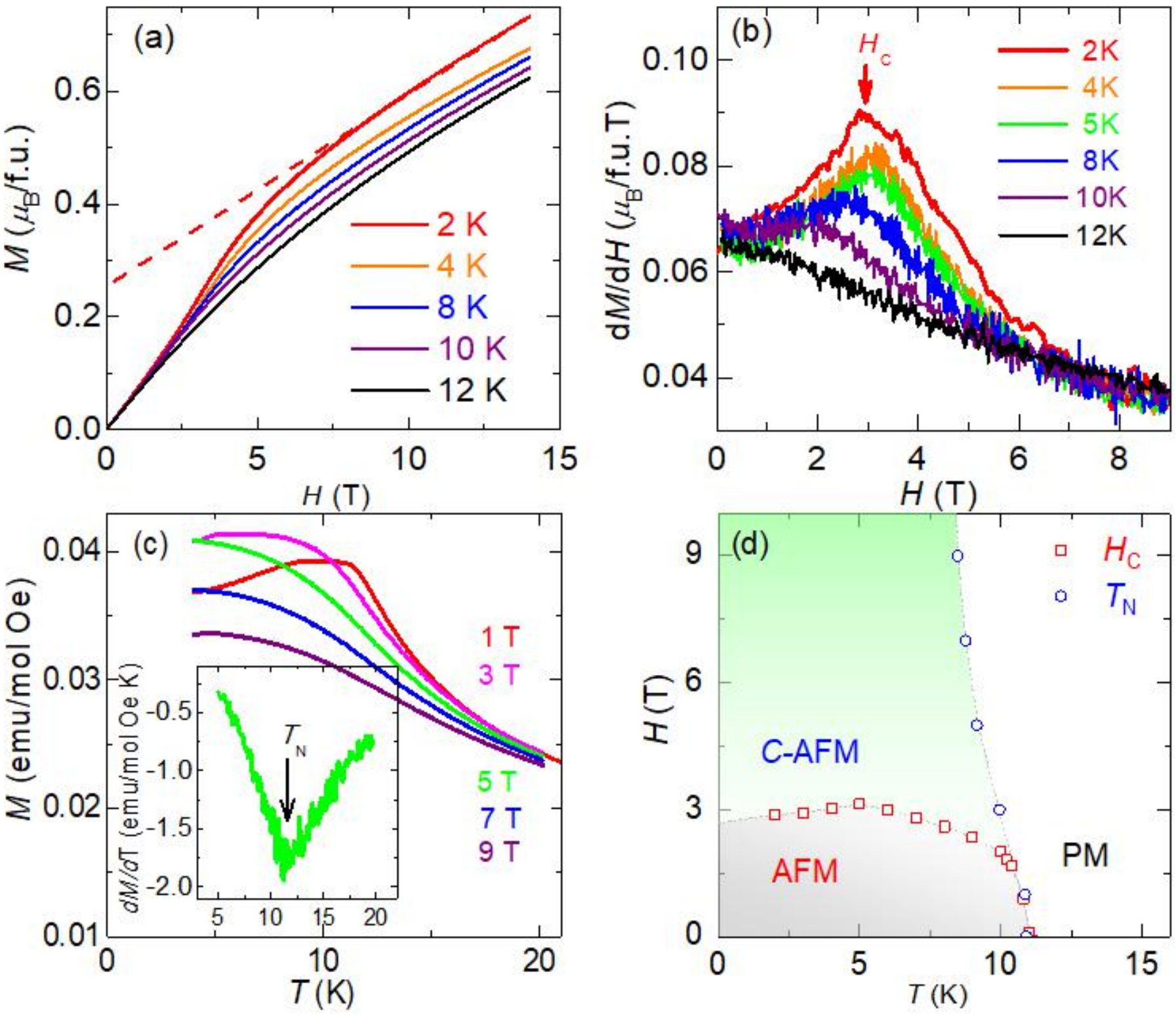

Figure 6

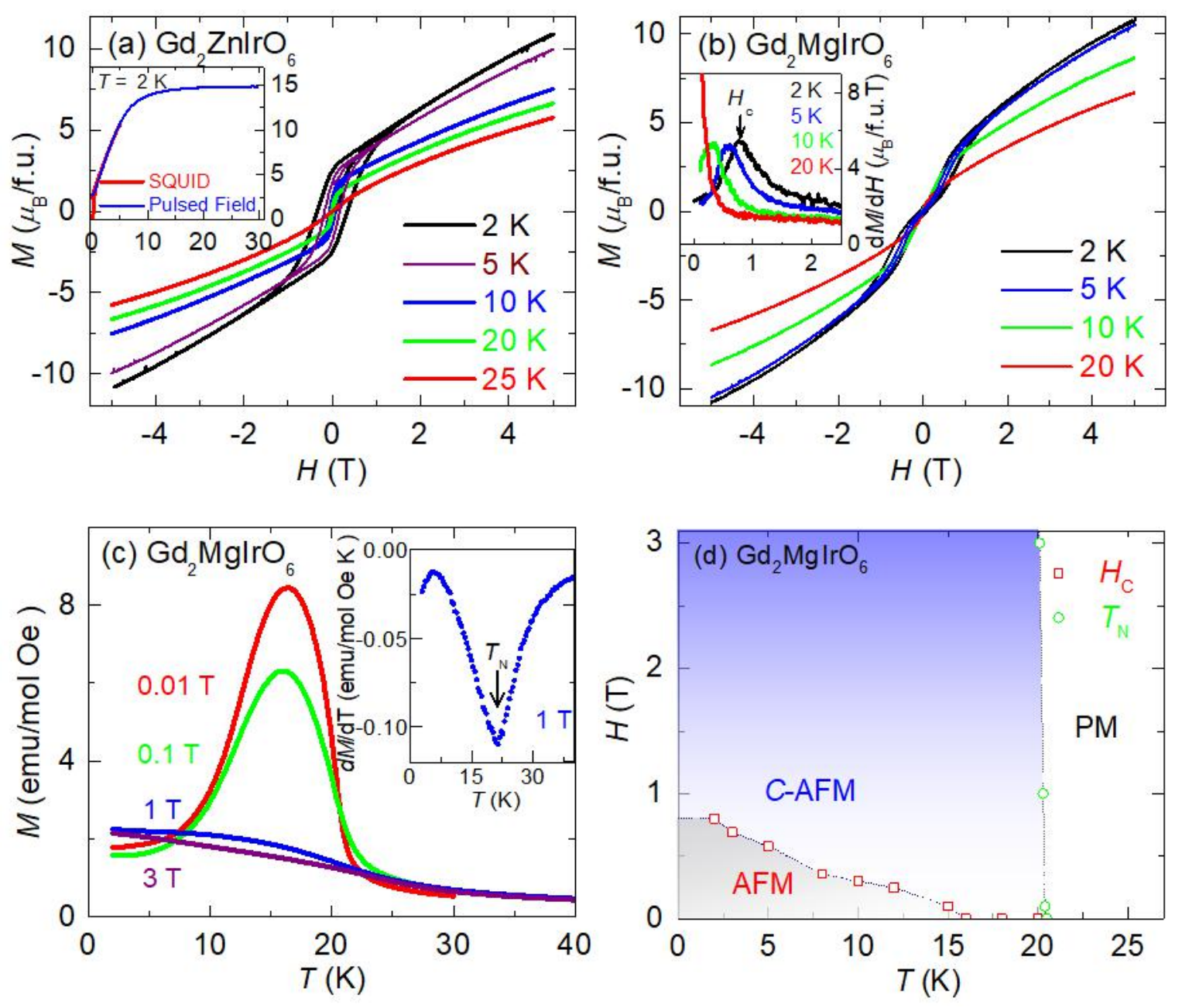

Figure 7

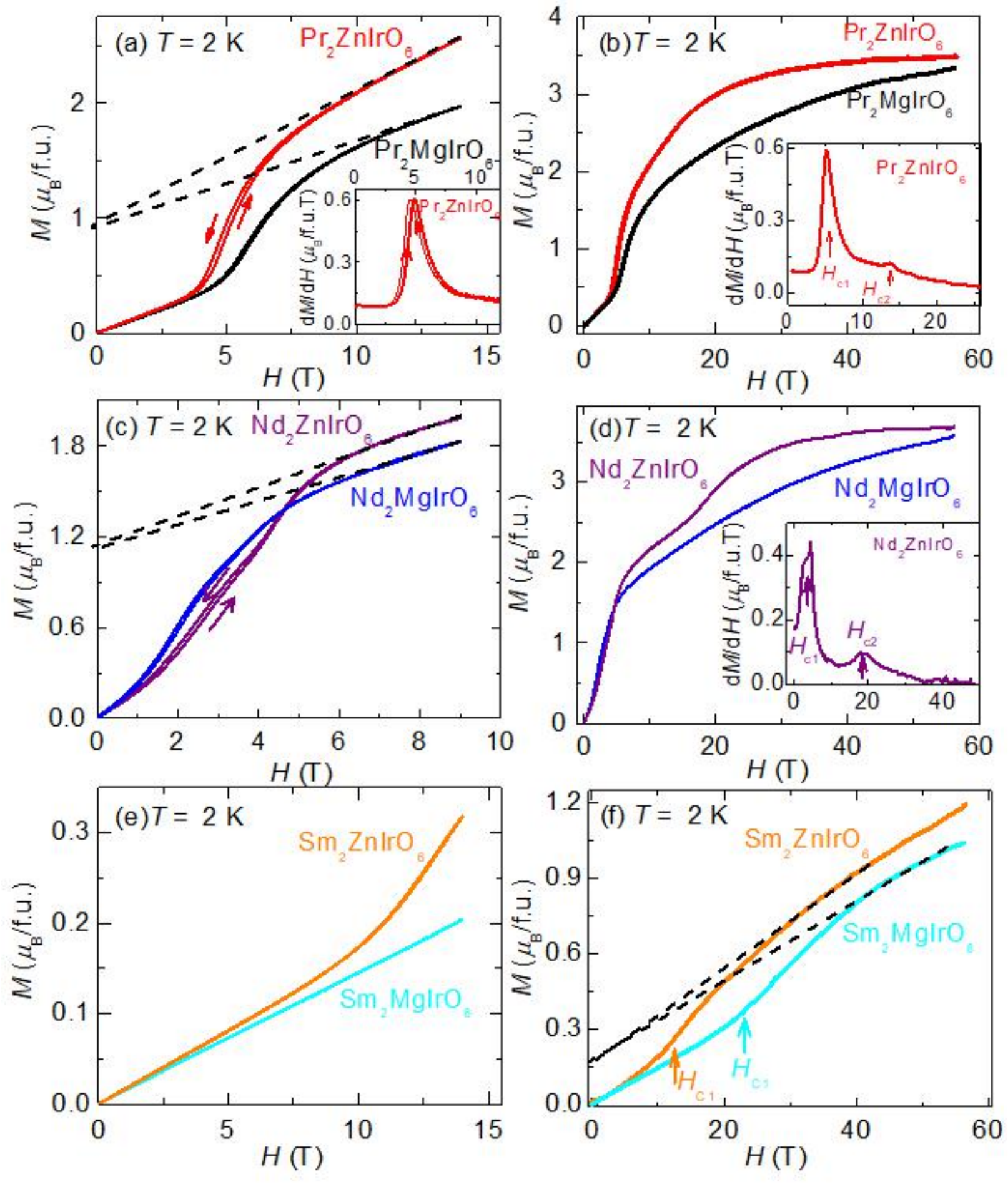

Figure 8

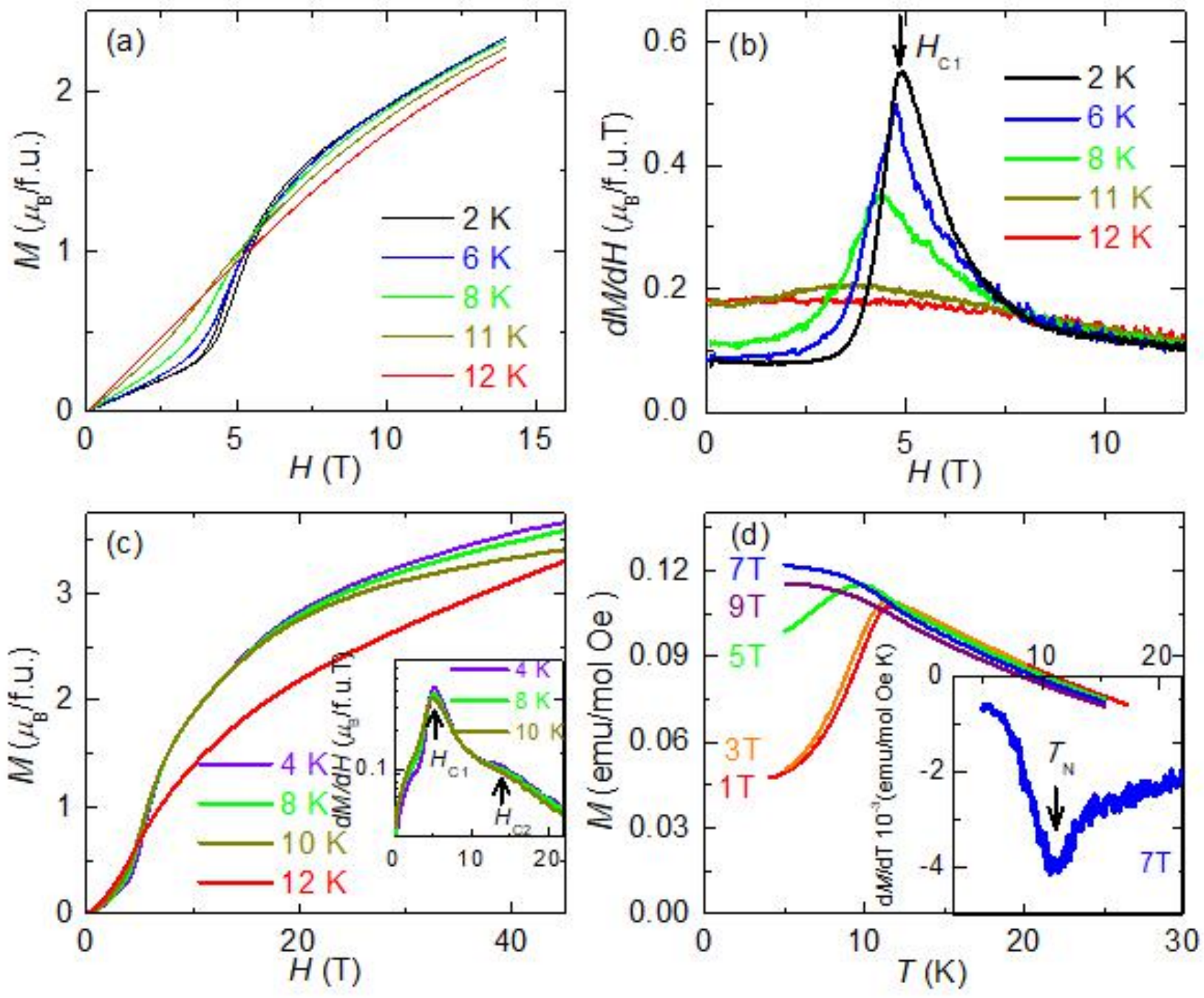

Figure 9

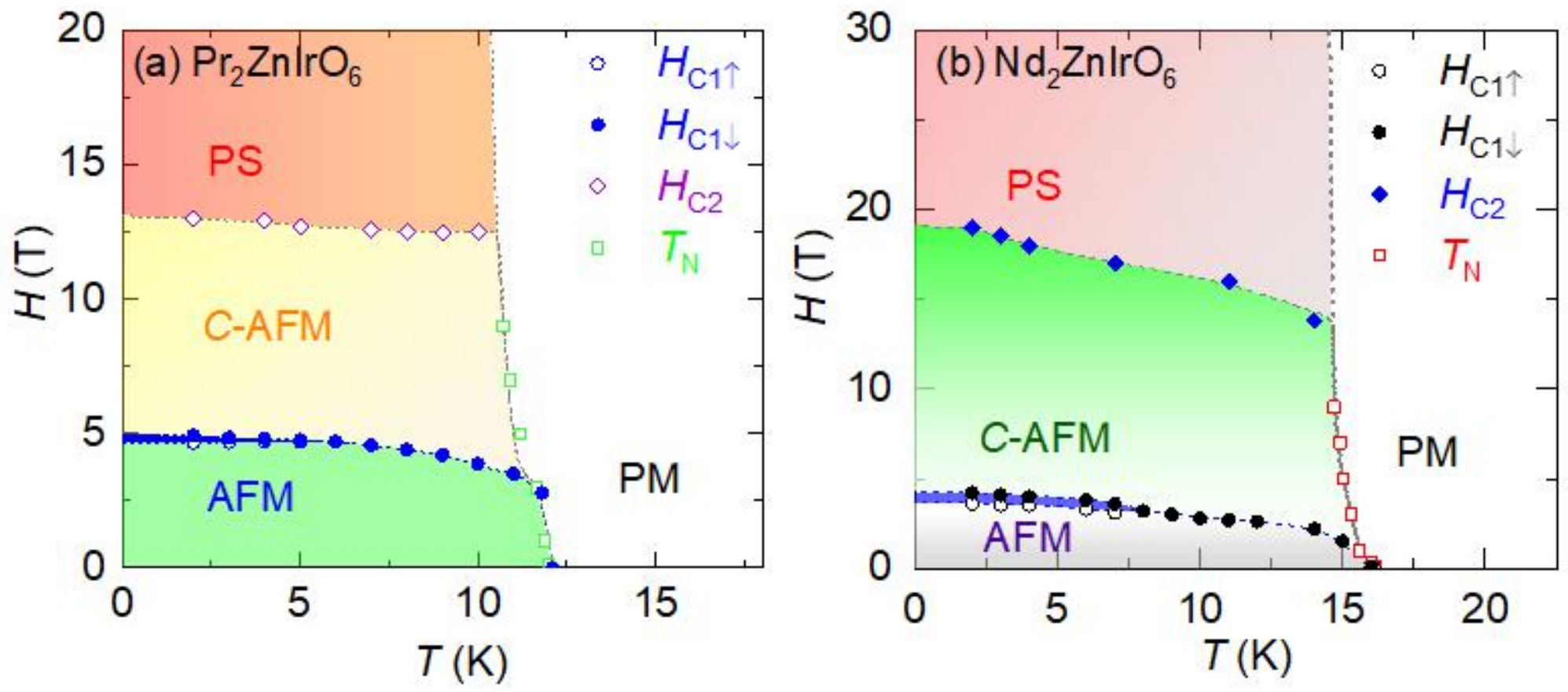

Figure 10